\documentclass[reprint, amsmath, amssymb]{revtex4-2}

\makeatletter
\def\@process@society#1{}
\makeatother

\usepackage[english]{babel}
\usepackage{dsfont}
\usepackage{graphicx}
\usepackage{dcolumn}
\usepackage{bm}
\usepackage[colorlinks,linkcolor=blue,anchorcolor=blue,citecolor=blue]{hyperref}

\begin{document}

	\title{Grey-body Factors and Absorption Cross Sections of Non-Commutative Black Holes under Einstein-Coupled Scalar Fields}
	
	\author{Sihao Fan}
	\email{sihaofan02@gmail.com}
	\affiliation{
		University of Shanghai for Science and Technology, Shanghai 200093, China
	}
	
	\author{Chen Wu}
	\email{wuchenoffd@gmail.com}
	\affiliation{Xingzhi College, Zhejiang Normal University, Jinhua 321004, Zhejiang, China}
    \author{Wenjun Guo}
	\affiliation{
		University of Shanghai for Science and Technology, Shanghai 200093, China
	}
	\date{\today}
	
	\begin{abstract}
	    This paper investigates scalar field perturbations coupled to the Einstein tensor of non-commutative black holes. We compute the grey-body factors and absorption cross-sections for different choices of the parameters using the partial wave method, and verify the latest correspondence between grey-body factors and quasinormal modes. The results show that larger values of the non-commutativity parameter $\theta_{NC}$ and the coupling constant $\eta$ introduced in this model lead to smaller absorption cross-sections. Furthermore, we find that this correspondence is accurate for non-commutative black holes in the limit of large angular momentum quantum number $l$.
	\end{abstract}
	
	\maketitle
	
	\section{Introduction} \label{sec1}
    Within the framework of black hole physics, the grey-body factor and quasinormal modes are two fundamental observables. Quasinormal modes describe the characteristic oscillation patterns of a black hole after perturbations. Their corresponding complex frequencies are determined by specific boundary conditions: purely outgoing waves at spatial infinity and purely ingoing waves at the event horizon respectively. The grey-body factor, on the other hand, is defined as the transmission coefficient of a field across the black hole's potential barrier. It characterizes the transmissivity of a field propagating from the event horizon to spatial infinity, where the boundary conditions require a pure ingoing wave at the event horizon and a superposition of ingoing and outgoing waves at infinity. As a crucial parameter for understanding black hole absorption and scattering behavior, the grey-body factor is widely employed in analyses of Hawking radiation processes \cite{hawking1975particle,page1976particle,page1976particle2}. Compared to the ideal black-body model, the grey-body factor provides a more realistic description of the propagation behavior of various fields (such as electromagnetic fields and gravitational waves) across the black hole's potential barrier \cite{konoplya2025correspondence}.
	
	In black hole physics, there exists a close relationship between the grey-body factor and the absorption cross-section. Extensive research has been devoted to the study of the absorption and scattering characteristics of massless scalar fields in various black hole backgrounds. Significant work has been conducted on static black holes \cite{abdalla2019instability,vieira2016confluent,wan_absorption_2023} and rotating black holes \cite{abdujabbarov2015energetics,leite2016absorption,leite2017scalar}. However, the scattering dynamics and related properties of various novel black hole models remain to be comprehensively investigated.
	
	This paper investigates non-commutative black holes. Several approaches are available for constructing gravitational theories based on general relativity in non-commutative spacetime. For example, one approach involves deforming Einstein gravity via the Seiberg-Witten map, utilizing the non-commutative SO(4,1) de Sitter group and its contraction to the ISO(3,1) group \cite{chamseddine2001deforming}. Alternatively, theories built upon the twisted Poincaré algebra have also been proposed \cite{aschieri2006noncommutative,kobakhidze2008theta}. In this work, we adopt a formulation of non-commutative gravity that is motivated by symmetry considerations, specifically using $\theta^{\mu\nu} = \theta \, \text{diag}(\varepsilon_{ij}, \varepsilon_{ij}, \ldots)$ \cite{smailagic2004lorentz}. This approach assists in resolving issues related to Lorentz invariance and unitarity.
	
	The study of non-commutative black holes has significant physical implications. Firstly, it offers a potential way to circumvent difficulties encountered in quantum field theory on curved spacetime. Secondly, non-commutativity can eliminate the central gravitational singularity of black holes \cite{nicolini2009noncommutative,araujo_filho_particle_2025,filho_effects_2024}. Additionally, non-commutativity has been shown to resolve other issues in standard black hole models, such as temperature divergence and unbounded curvature growth \cite{gruppuso2005newton,nicolini2005model,filho_axisymmetric_2025}. As the spacetime of a non-commutative black hole is not a vacuum solution to Einstein's equations, the Einstein tensor $G_{\mu\nu}$ contributes to the system dynamics even under spherical symmetry. Inspired by this, we investigate the physical behavior of scalar fields coupled to the Einstein tensor within this non-commutative gravitational background.
	
	A correspondence between quasinormal modes and grey-body factors has been identified, due to the fact that both quantities are derived from the same underlying wave equation, despite their different boundary conditions and physical interpretations. Studies have revealed that an explicit analytical relation connects them in the high-frequency regime \cite{okabayashi2024greybody,oshita2023greybody,oshita2023thermal}. In the case of spherically symmetric and asymptotically flat black holes, a correspondence between quasinormal modes and grey-body factors has been established in the eikonal limit \cite{konoplya2024correspondence}. { Moreover, this correspondence has been extended to axisymmetric black holes \cite{konoplya2025correspondence} and wormholes \cite{bolokhov2025correspondence}, and its accuracy has been further verified across a broad range of gravitational backgrounds, including Einstein–Euler–Heisenberg (EEH) theory \cite{malik_grey-body_2025}, asymptotically de Sitter spacetimes \cite{malik2025correspondence}, and black holes in modified gravity or quantum-corrected spacetimes \cite{lutfuoglu2025black,lutfuoglu2026long}.}The theoretical foundation for this correspondence is provided by the WKB approximation method, which produces highly accurate results under the eikonal limit. {\color{red} Compared to quasinormal modes, grey-body factors exhibit remarkable robustness against small deformations of the background spacetime. Konoplya \& Pappas \cite{konoplya_dirty_2025} demonstrated that deformations induced by near-horizon quantum effects or far-field astrophysical environments only lead to weak corrections (with maximum deviations below 1\%) to low-frequency grey-body factors, whereas the frequencies and damping rates of high-overtone quasinormal modes can deviate by tens of percent. Rosato et al. \cite{rosato_ringdown_2024} further verified by introducing Pöschl–Teller perturbations to the Schwarzschild effective potential that, even when the quasinormal mode spectrum is strongly distorted, grey-body factor deviations remain at linear order $O(\epsilon)$ and decrease as the perturbation moves away from the potential peak. The grey-body factor thus maintains stability and serves as a reliable characterization of the high-frequency amplitude in gravitational-wave ringdown.}
	
	Despite its utility, the WKB approximation is not applicable universally. Its validity, for example, can break down in theories incorporating higher-order curvature corrections \cite{aoki2020consistent,konoplya2017eikonal}. {\color{red}Rather than testing the existence of the correspondence itself, which is already well understood in the eikonal regime, the goal of this work is to investigate how it is affected by non-commutativity and non-minimal scalar coupling. In this work, we employ the sixth-order WKB approximation  \cite{iyer1987black,konoplya2003quasinormal,schutz1985black} as a semi-analytical framework to evaluate quasinormal modes and, through the established correspondence, infer the grey-body factors and absorption cross sections. }Previous studies indicate that computing quasinormal modes in non-commutative spacetime using the WKB method introduces certain errors \cite{das2019quasinormal,yan2021quasinormal}. However, for small values of the non-commutative parameter $\theta_{NC}$ or the coupling parameter $\eta$, the accuracy remains acceptable \cite{yan2021quasinormal,zhou_non-minimal_2013,lin_dynamical_2011}. Furthermore, studies have identified scenarios where WKB-based methods prove to be inapplicable \cite{konoplya2019higher}. {\color{red}We therefore interpret the agreement with first-order WKB results in the large-$\ell$ regime as a consistency check of the correspondence rather than an independent verification.}
	
	This study investigates the properties of black holes in non-commutative spacetime. Specifically, we calculate the grey-body factors and absorption cross-sections for scalar fields non-minimally coupled to the Einstein tensor. We systematically examine how key parameters—namely, the non-commutative parameter $\theta_{NC}$ and the coupling constant $\eta$—affect these scattering observables. Our objective is to clarify the unique wave-propagation characteristics of these black holes, which incorporate quantum-gravity corrections.
	
	\section{Wave Equation with Einstein Tensor Coupling} \label{sec2}
	
	The line element for a four-dimensional static spherically symmetric black hole spacetime can be written as:
	\begin{equation}
		\label{eq:metric}
		ds^2 = -f(r)\,dt^2 + \frac{1}{f(r)}\,dr^2 + r^2 d\Omega^2,
	\end{equation}
	where $d\Omega^2 = d\theta^2 + \sin^2\theta \, d\phi^2$ is the metric on the two-sphere.
	
	This work considers a massive scalar field coupled to the Einstein tensor \cite{amendola1993cosmology,sushkov2009exact} :
	\begin{equation}
		\label{eq:lagrangian}
		\mathcal{L}_{\mathrm{pert}} = -\frac{\sqrt{-g}}{2} \left[ (g^{\mu\nu} + \eta G^{\mu\nu}) \, \partial_\mu \Phi \, \partial_\nu \Phi + m^2 \Phi^2 \right],
	\end{equation}
	where $\eta$ is the non-minimally derivative coupling parameter. The magnitude of $\eta$ governs the strength of the coupling to the Einstein tensor, and $m$ is the mass of the scalar field.
	
	The equation of motion for the scalar field is given by:
	\begin{equation}
		\label{eq:eom}
		\frac{1}{\sqrt{-g}} \partial_\mu \left[ \sqrt{-g} (g^{\mu\nu} + \eta G^{\mu\nu}) \partial_\nu \Psi \right] - m^2 \Psi = 0.
	\end{equation}
	
	Performing separation of variables on the scalar field, we employ the ansatz:
	\begin{equation}
		\label{eq:ansatz}
		\Psi(t, r, \theta, \phi) = \sum_{l, m} e^{-i\omega t} \frac{b(r)}{r} R(r) \, Y_{l,m}(\theta, \phi),
	\end{equation}
	where $\omega$ is the complex frequency, $Y_{l,m}(\theta, \phi)$ are the spherical harmonics, and the factor $b(r)/r$ is introduced to simplify the resulting radial equation. The function $b(r)$ may incorporate factors related to the specific coupling, such as $b(r) = (1 - \eta F(r))$ in some models.
	
	The spherically symmetric Einstein tensor components $G^{\mu\nu}$ are given by:
	\begin{equation}
		G^{\mu\nu} = \begin{pmatrix}
			\frac{A(r)}{f(r)} & 0 & 0 & 0 \\
			0 & -A(r) f(r) & 0 & 0 \\
			0 & 0 & \frac{B(r)}{r^2} & 0 \\
			0 & 0 & 0 & \frac{B(r)}{r^2 \sin^2\theta}
		\end{pmatrix},
	\end{equation}
	where the functions $A(r)$ and $B(r)$ are defined as:
	\begin{align}
		A(r) &= \frac{1 - f(r)}{r^2} - \frac{f'(r)}{r}, \\
		B(r) &= A(r) - \frac{\mathcal{R}}{2} = \frac{f'(r)}{r} + \frac{f''(r)}{2}.
	\end{align}
	Here, the Ricci scalar $\mathcal{R}$ is:
	\begin{equation}
		\mathcal{R} \equiv -\left[ f''(r) + \frac{4 f'(r)}{r} + \frac{2(f(r)-1)}{r^2} \right].
	\end{equation}
	We also define the function $b(r)$ as:
	\begin{equation}
		b(r) \equiv \frac{1}{r \sqrt{1 - \eta A(r)}}.
	\end{equation}
	
    The value of the coupling parameter $\eta$ must ensure the positive definiteness of the effective kinetic term to avoid unphysical phenomena such as superluminal propagation or imaginary sound speeds in the scalar field \cite{gao_when_2010}. Studies on various black hole backgrounds, including Schwarzschild--de Sitter black holes \cite{zhou_non-minimal_2013} and Reissner--Nordstr\"{o}m black holes \cite{lin_dynamical_2011}, have systematically analyzed the effective potential structure and quasinormal mode properties of the Einstein tensor-coupled scalar field, verifying the physical self-consistency of this coupling within reasonable parameter ranges. In particular, the study presented in Ref.\cite{fontana_dynamical_2019} explicitly provided a quantitative constraint for the non-minimal derivative coupling: $\eta < L^2/3$, where $L$ is the de Sitter radius. This paper adopts the same parameter selection principle as the aforementioned literature, strictly satisfying the conditions for positive definiteness of the kinetic term and causal propagation, thereby effectively avoiding issues of dynamical instability.

	In the tortoise coordinate $dr_* = dr / f(r)$, expanding and rearranging the equation of motion yields a Schr\"{o}dinger-like equation:
	\begin{equation}
		\label{eq:schrodinger_eq}
		\frac{\partial^2 R}{\partial r_*^2} + \left[ \omega^2 - V(r) \right] R = 0,
	\end{equation}
	where the effective potential $V(r)$ is given by:
	\begin{equation}
		\label{eq:effective_potential}
		\small
		\begin{split}
			V_{\text{eff}}(r) = &\frac{f(r)}{1 - \eta A(r)} \left[ \frac{l(l+1)}{r^2} (1 + \eta B(r)) + m^2 \right] \\
			&+ f(r)^2 \left[ \frac{-\eta}{1 - \eta A(r)} \left( \frac{A''(r)}{2} + \frac{A'(r) f'(r)}{2f(r)} + \frac{A'(r)}{r} \right) \right. \\
			&\left. + \frac{f'(r)}{r f(r)} - \frac{1}{4} \left( \frac{-\eta A'(r)}{1 - \eta A(r)} \right)^2 \right].
		\end{split}
	\end{equation}
	The metric function used in this work originates from non-commutative geometry \cite{nicolini2006noncommutative} :
	\begin{equation}
		\label{eq:metric_function}
		f(r) = 1 - \frac{4M}{r\sqrt{\pi}} \gamma\left(\frac{3}{2}, \frac{r^2}{4\theta_{NC}}\right),
	\end{equation}
	where $\gamma$ is the lower incomplete gamma function. An equivalent form is given by:
	\begin{equation}
		\label{eq:metric_equivalent}
		f(r) = 1 - \frac{M r^2}{3\sqrt{\pi} \theta_{NC}^{3/2}} \mathcal{M}\left(\frac{3}{2}, \frac{5}{2}, -\frac{r^2}{4\theta_{NC}}\right).
	\end{equation}
	
	This metric reduces to the Schwarzschild solution in the limit $r^2 \gg 4\theta_{NC}$. Here, $\mathcal{M}$ denotes the confluent hypergeometric function, and this particular form of the metric has been shown to yield favorable results in practical computations \cite{yan2021quasinormal}.
	
	Based on the properties of the effective scattering potential, the wave equation can be solved. Its asymptotic solutions are:
	
	\begin{equation}
		\label{eq:asymptotic_solution}
		\resizebox{\linewidth}{!}{$
			\Psi_{\omega l} \approx
			\begin{cases}
				T_{\omega l} \omega e^{-i\omega\chi}, & \chi \to -\infty; \\
				\omega\chi \left[ (-i)^{l+1} h_l^{(1)*}(\omega\chi) + i^{l+1} R_{\omega l}(\omega) h_l^{(1)}(\omega\chi) \right], & \chi \to +\infty;
			\end{cases}
		$}
	\end{equation}
	
	where $h_l^{(1)}(\omega\chi)$ is the spherical Hankel function of the first kind \cite{abramowitz1964handbook}. {{Here, $\chi$ is defined in tortoise coordinates as:$\chi = \int f^{-1}(r)\,dr.$}} The quantities $|T_{\omega l}|^2$ and $|R_{\omega l}|^2$ represent the transmission and reflection coefficients, respectively, and satisfy the probability conservation equation:
	\begin{equation}
		\label{eq:probability_conservation}
		|T_{\omega l}|^2 + |R_{\omega l}|^2 = 1.
	\end{equation}
	
	Considering the asymptotic form of the spherical Hankel function for $|\omega \chi| \gg l(l+1)/2$:
	\begin{equation}
		h_l^{(1)}(\omega\chi) = j_l(\omega\chi) + i n_l(\omega\chi) \approx (-i)^{l+1} \frac{e^{i \omega \chi}}{\omega \chi},
	\end{equation}
	where $j_l$ and $n_l$ are the spherical Bessel and Neumann functions, Eq.~(\ref{eq:asymptotic_solution}) can be approximated as:
	\begin{equation}
		\label{eq:approx_solution}
		\Psi_{\omega l} \approx
		\begin{cases}
			A_{\omega l}^{tr} e^{-i \omega \chi}, & \chi \to -\infty; \\
			A_{\omega l}^{in} e^{-i \omega \chi} + A_{\omega l}^{out} e^{+i \omega \chi}, & \chi \to +\infty;
		\end{cases}
	\end{equation}
	where the coefficients $A_{\omega l}^{tr}(\omega)$, $A_{\omega l}^{in}(\omega)$, and $A_{\omega l}^{out}(\omega)$ are related to the incident, reflected, and transmitted partial wave amplitudes, respectively.

	Consequently, the grey-body factor is defined as:
	\begin{equation}
		\label{eq:greybody_definition}
		\Gamma_l(\omega) = 1 - \left| \frac{A_{\omega l}^{out}}{A_{\omega l}^{in}} \right|^2.
	\end{equation}

     {\color{red} For $\omega > 0$ and $l \in \mathbb{N}$, Iyer et al.\cite{iyer1987black,decanini2010unstable,decanini2011fine} derived the expression for the grey-body factor as:}
     	
     	\begin{equation}
     			\Gamma_{\omega l}(\omega) = \frac{1}{1 + \exp\left[-2\pi \frac{\omega^2 - V_l(x_p)}{\sqrt{-2V_l''(x_p)}}\right]},
     	\end{equation}
    
     where $x_p$ denotes the value corresponding to the maximum of the effective potential in polar coordinates.

    \begin{figure*}
        \includegraphics[width=0.9\textwidth]{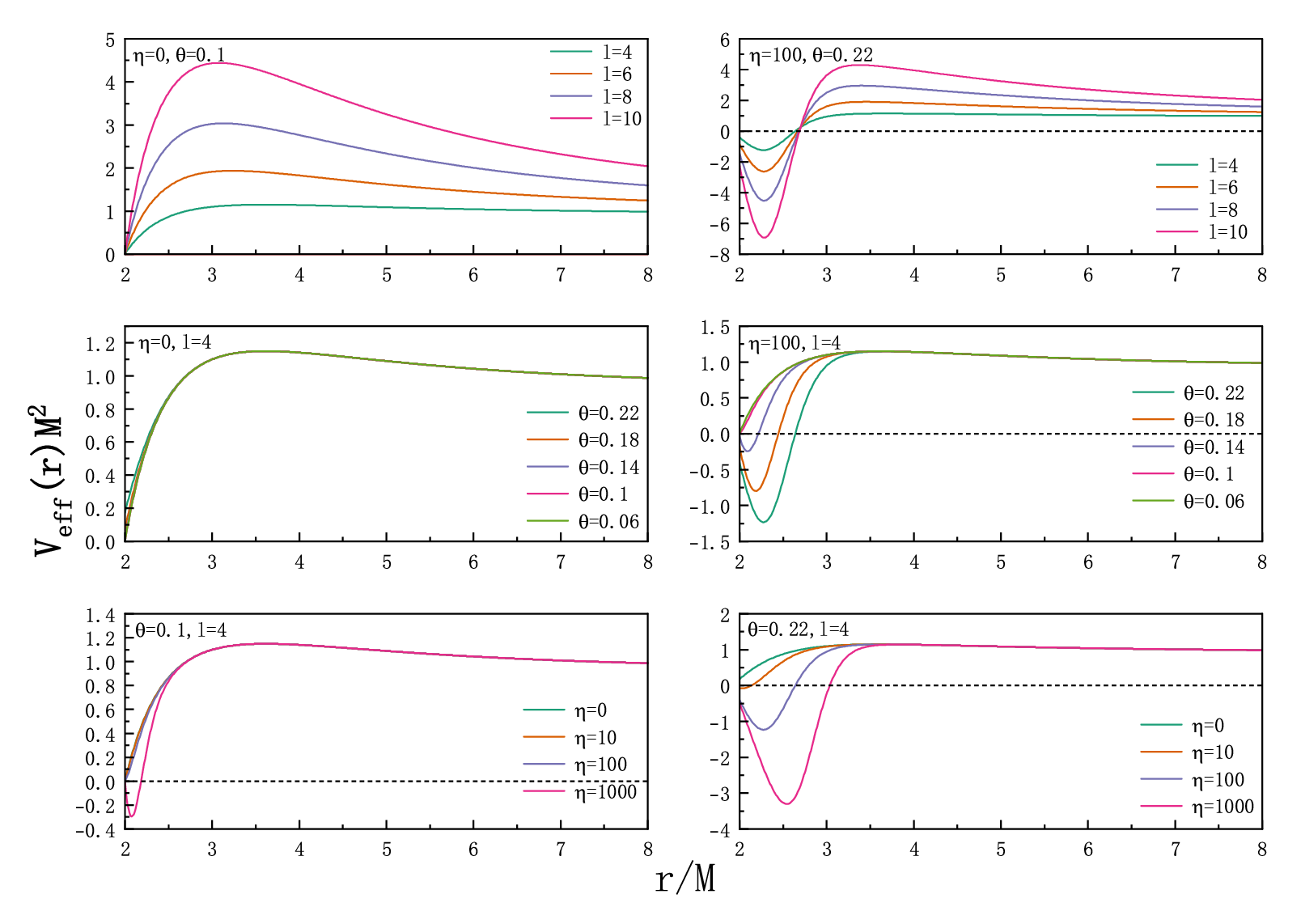}
        \caption{\label{fig1}Effective potential energy diagrams for different values of {$l$}, {$\theta_{NC}$}and {$\eta$},with $M$=1.}
    \end{figure*}

    \begin{figure*}
	  \includegraphics[width=0.5\textwidth]{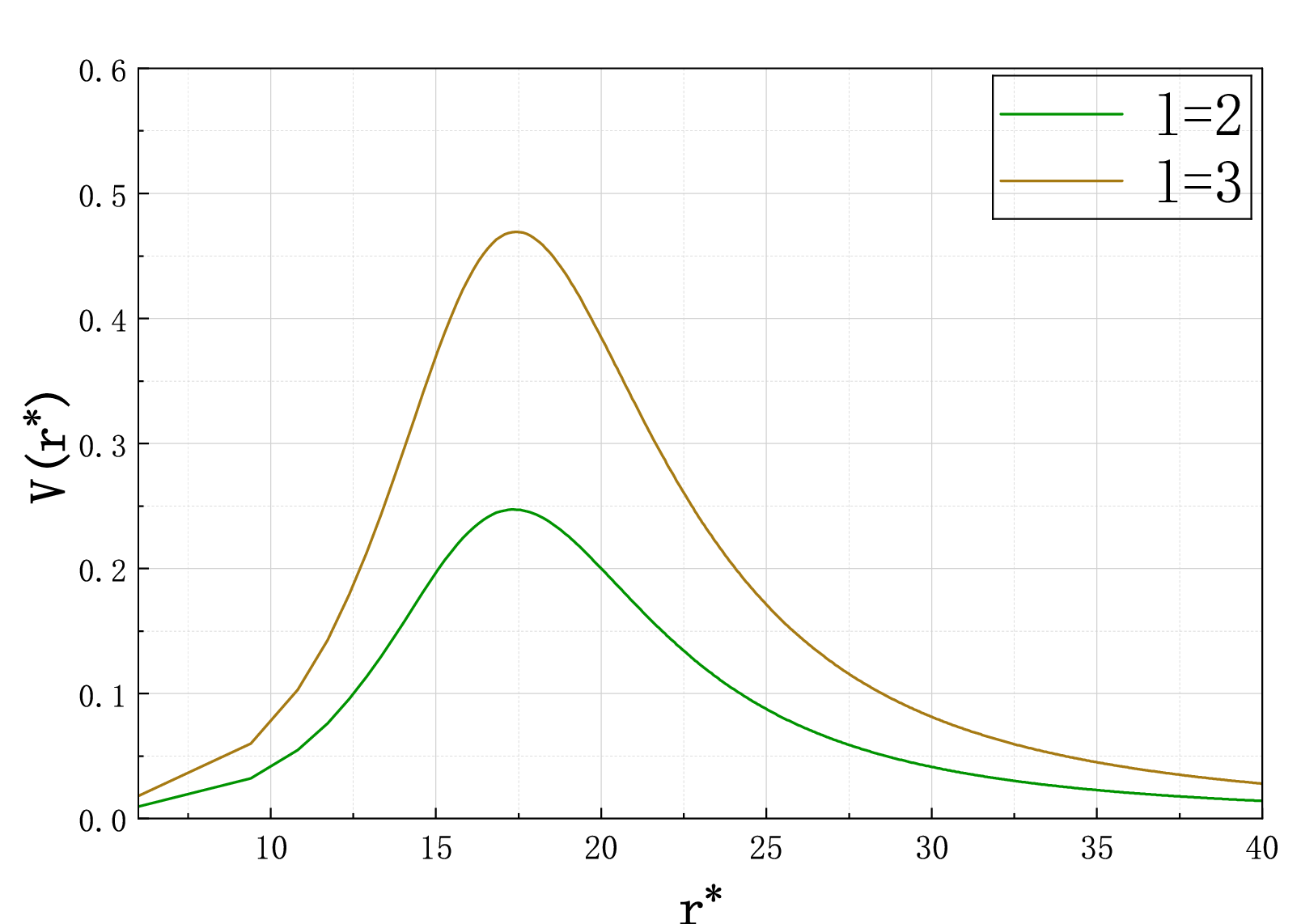}
	  \caption{\label{fig2}Effective potential in the tortoise coordinate for $\theta_{NC} = 0.2$, $\eta = 100$, and $\ell = 2, 3$.}
    \end{figure*}

    Figure \ref{fig1} displays the effective potential in spherical coordinates for various parameters \( l \), \( \eta \), and \( \theta_{NC} \), where \( l \) is the angular momentum quantum number, \( \eta \) is the nonminimally derivative coupling parameter, and \( \theta_{NC} \) is the noncommutative parameter. {Figure \ref{fig2} shows the effective potential in the tortoise coordinate for $\theta = 0.2$, $\eta = 100$, and $\ell = 2, 3$. The potential exhibits a shape analogous to a standard quantum-mechanical scattering barrier.} 
    
    Following are the observations: With fixed \( \eta \) and \( \theta_{NC} \), a larger \( l \) leads to a higher effective potential and a sharper peak. Moreover, when both \( \eta \) and \( \theta_{NC} \) are relatively large compared to when they are small, the negative potential region exhibits significant differences. Notably, the value of \( l \) does not alter the position of the potential zero point.
    
    The parameter \( \theta_{NC} \), which governs the non-commutative properties of the metric, falls within a valid range \( 0 < \theta_{NC} < 0.275811 \) \cite{nicolini2006noncommutative,yan2020scalar}. A larger \( \theta_{NC} \) also leads to an expansion of the negative potential region, and the magnitude of \( \eta \) influences how \( \theta_{NC} \) affects the effective potential. A larger \( \eta \) causes the variation of \( \theta_{NC} \) to have a more pronounced impact on the effective potential. As defined by the metric function in equation (\ref{eq:metric_function}), at large \( r \), the metric reverts to the Schwarzschild metric, so \( \theta_{NC} \) exerts a weaker influence at large \( r \). Additionally, a larger \( \theta_{NC} \) displaces the zero point of the potential to a greater radial distance.
	
	 For fixed \( \theta_{NC} \) and \( l \), increasing \( \eta \) leads to an expansion of the negative potential region and a sharper minimum peak. The parameter \( \eta \) represents the strength of the coupling between the scalar field and gravity—a larger \( \eta \) is associated with a stronger gravitational field. This expanded negative region indicates that particles experience a stronger gravitational pull, similar to entering a potential well. Similarly, a larger \( \theta_{NC} \) enhances the effect of \( \eta \) on the effective potential. Consistent with the behavior of gravitational fields, the influence of \( \eta \) decreases at large \( r \). Additionally, a larger \( \eta \) moves the potential's zero point to a larger radial distance.

	\section{Quasinormal Modes and Grey-body Factors} \label{sec3}
	Recently, it has been shown that the grey-body factor can be calculated using the known fundamental mode and the first overtone \cite{konoplya2024correspondence}. This formula is exact in the limit $l\rightarrow\infty$ and provides an approximation for small $l$.
	
	The WKB expression for the grey-body factor is:
	\begin{equation}
		\Gamma_l\left(\omega\right)=\frac{1}{1+e^{2\pi i\mathcal{K}}},
	\end{equation}
	where $\mathcal{K}$ satisfies the equation:
    \begin{equation}\label{eq20}
    	\small
    	\begin{split}
		  \omega^2 &= V_0 + A_2(\mathcal{K}^2) + A_4(\mathcal{K}^2) + A_6(\mathcal{K}^2) \\
		  &\quad \pm i\mathcal{K}\sqrt{-2V_2}\left(1 + A_3(\mathcal{K}^2) + A_5(\mathcal{K}^2) + A_7(\mathcal{K}^2)\ldots\right).
    	\end{split}
    \end{equation}
	Here, $V_0$ is the maximum value of the effective potential, $V_2$ is the second derivative of the effective potential at its maximum, and $A_i$ for $i=2,3,4,\ldots$ are the $i$-th order WKB correction terms beyond the eikonal approximation. These terms depend on the derivatives of the potential at its maximum up to the order $2i$. The explicit forms of $A_i$ for the second and third WKB orders were given by \cite{iyer1987black}, for the fourth to sixth orders by \cite{konoplya2003quasinormal}, and for the seventh to thirteenth orders by \cite{matyjasek2017quasinormal}.
	
	where $\mathcal{K}$ satisfies:
	\begin{equation}
		\mathcal{K} = n + 1/2, \quad n = 0, 1, 2 \ldots
	\end{equation}
	Here, $n$ is the overtone number: when $n=0$, formula (\ref{eq20}) gives the dominant quasinormal frequency $\omega_0$, which corresponds to the mode with the slowest decay rate.
	
	Research provided the correspondence between the grey-body factor and quasinormal modes for large $l$ \cite{konoplya2024correspondence}.
	In the eikonal approximation, we have:
	\begin{equation}
		i\mathcal{K} = \frac{\omega^2 - \text{Re}(\omega_0)^2}{4\text{Re}(\omega_0)\text{Im}(\omega_0)} + \mathcal{O}\left(\frac{1}{l}\right)
	\end{equation}
	
	{The value of $\mathcal{K}$ taking into account the first-order beyond-eikonal correction is:}
	
	\begin{equation}
		i\mathcal{K} = \frac{\omega^2 - {\text{Re}(\omega_0)}^2}{4\text{Re}(\omega_0)\text{Im}(\omega_0)} - \frac{\text{Re}(\omega_0) - \text{Re}(\omega_1)}{16\text{Im}(\omega_0)}
	\end{equation}
	where $\omega_1$ is calculated for $n=1$ ($\mathcal{K}=3/2$).
	
		{The value of $\mathcal{K}$ taking into account the second-order beyond-eikonal correction is:}
	
     \begin{equation}
     	\resizebox{0.5\textwidth}{!}{%
     		$\displaystyle
     		\begin{aligned}
     			i\mathcal{K} = &\frac{\omega^2 - {\text{Re}(\omega_0)}^2}{4\text{Re}(\omega_0)\text{Im}(\omega_0)} \left(1 + \frac{(\text{Re}(\omega_0) - \text{Re}(\omega_1))^2}{32{\text{Im}(\omega_0)}^2} - \frac{3\text{Im}(\omega_0) - \text{Im}(\omega_1)}{24\text{Im}(\omega_0)}\right) \\
     			&-\frac{\text{Re}(\omega_0) - \text{Re}(\omega_1)}{16\text{Im}(\omega_0)} - \frac{(\omega^2 - {\text{Re}(\omega_0)}^2)^2}{{16\text{Re}(\omega_0)}^3\text{Im}(\omega_0)} \left(1 + \frac{\text{Re}(\omega_0)(\text{Re}(\omega_0) - \text{Re}(\omega_1))}{{4\text{Im}(\omega_0)}^2}\right) \\
     			&+ \frac{(\omega^2 - {\text{Re}(\omega_0)}^2)^3}{{32\text{Re}(\omega_0)}^5\text{Im}(\omega_0)} \left(1 + \frac{\text{Re}(\omega_0)(\text{Re}(\omega_0) - \text{Re}(\omega_1))}{{4\text{Im}(\omega_0)}^2} + {\text{Re}(\omega_0)}^2 \times \right. \\
     			&\quad \left. \left( \frac{(\text{Re}(\omega_0) - \text{Re}(\omega_1))^2}{{16\text{Im}(\omega_0)}^4} - \frac{3\text{Im}(\omega_0) - \text{Im}(\omega_1)}{12\text{Im}(\omega_0)} \right) \right) + \mathcal{O}\left(\frac{1}{l^3}\right)
     		\end{aligned}
     		$%
     	}
     \end{equation}
	
\begin{figure*}
		\includegraphics[width=0.8\textwidth]{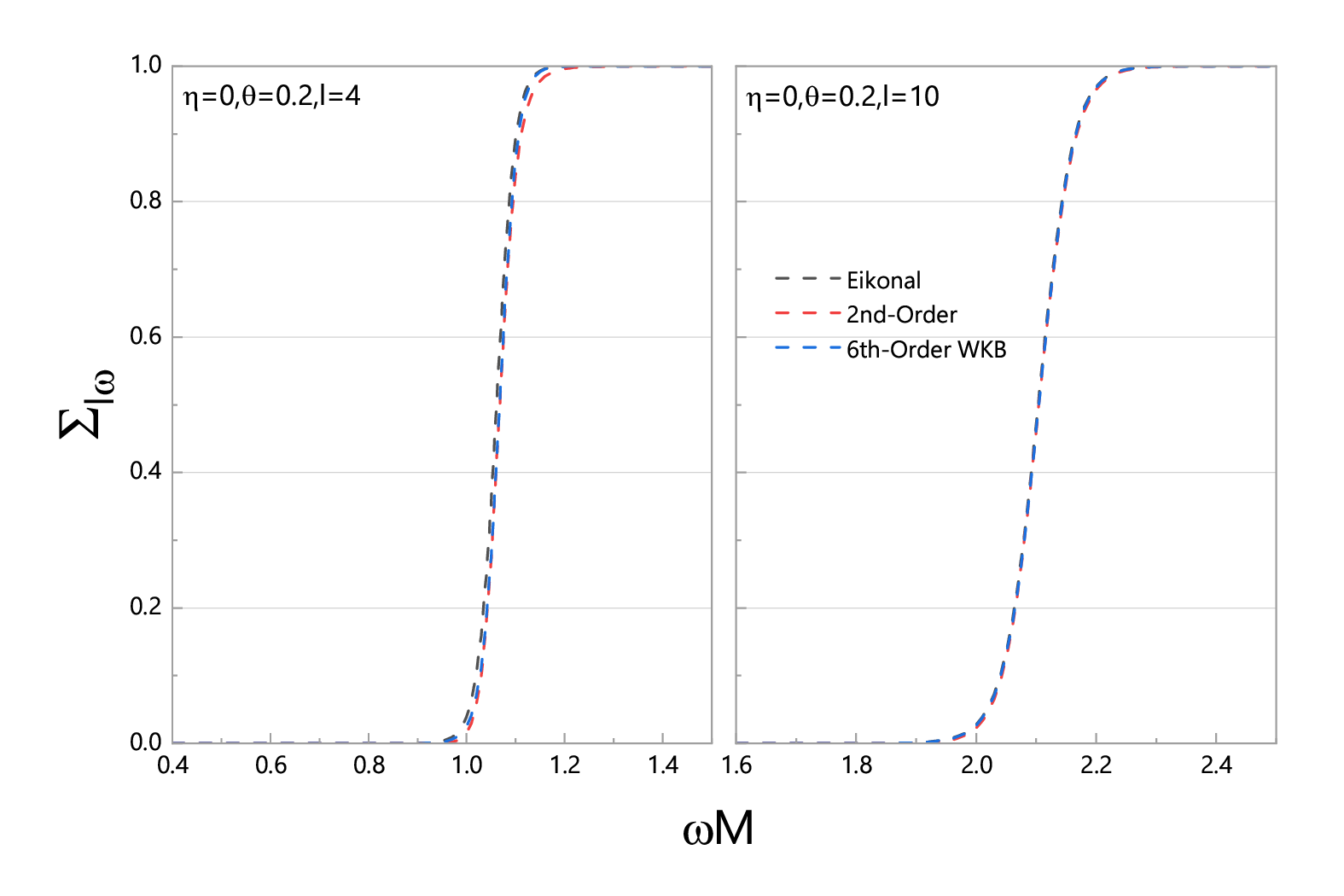}
		\caption{\label{fig3}The grey-body factors under different parameters.}
	\end{figure*}

    It has been pointed out that various WKB methods become inapplicable in several scenarios \cite{konoplya2019higher}. In general, the WKB approximation offers reasonable accuracy when $l > n$. We compare the grey-body factors obtained via the first-order (eikonal) WKB approach with those from the second-order and sixth-order WKB approximations. According to the results shown in Figure \ref{fig3}.
	
\begin{figure*}
		\includegraphics[width=0.9\textwidth]{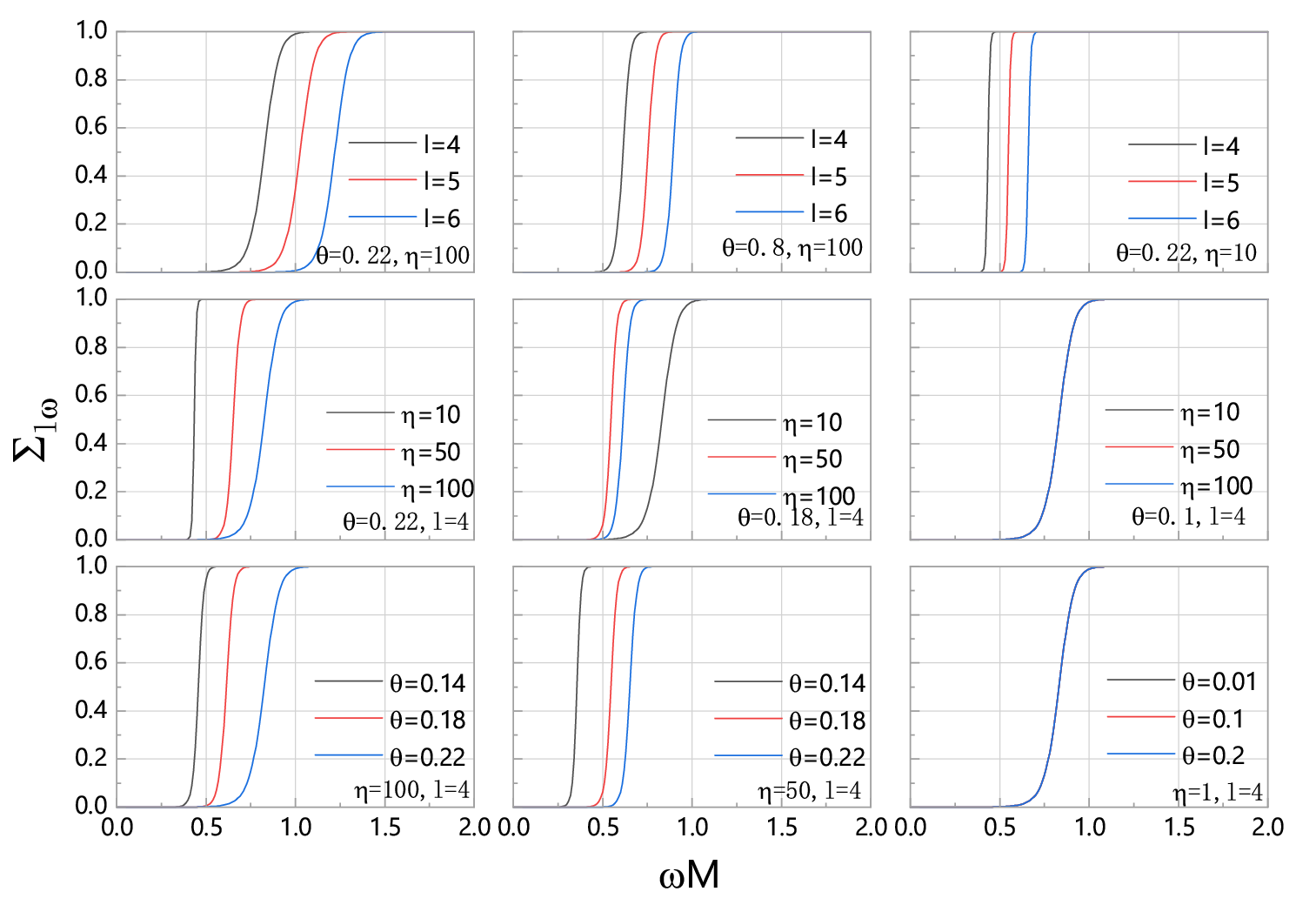}
		\caption{\label{fig4}Grey-body factors under different parameters.}
	\end{figure*}

	Figure \ref{fig4} displays the grey-body factors for different angular quantum numbers \( l \). A comparison reveals that the accuracy of the corresponding relationship remains relatively low for noncommutative black holes. However, as the angular quantum number \( l \) increases, the accuracy of the corresponding relationship improves significantly—including in the noncommutative case. This trend supports the inference that in the limit \( l \to \infty \), the corresponding relationship provides a highly accurate approximation for axisymmetric black holes \cite{konoplya2025correspondence}.
	
	 Figure \ref{fig4} depicts the dependence of the grey-body factors. As shown, the grey-body factor $\Gamma_{\omega l}$ approaches unity at large values of $\omega$ and tends to zero in the low-frequency range. As $l$ increases, the $\Gamma_{\omega l}$ curve shifts towards higher frequencies, which is consistent with the trend shown in Figure \ref{fig1} where a larger $l$ increases the peak height and the threshold frequency. Moreover, increasing either $\eta$ or $\theta_{NC}$ makes the grey-body factor curve flatter, leading to a more gradual transition from zero to unity. Similarly, an increase in $\theta_{NC}$ also leads to a more gradual variation in $\Gamma_{\omega l}$. Increasing either parameter also shifts the $\Gamma_{\omega l}$ curve towards lower frequencies. This phenomenon is associated with the effect of the parameters on the negative region of the effective potential. In addition, when either $\theta_{NC}$ or $\eta$ is relatively large, changes in the other parameter will significantly affect the grey-body factor.
	
    \section{Absorption cross section} \label{sec4}
	The grey-body factor is defined as the transmission coefficient: $T_{\omega l}(\omega)$:
	\begin{equation}
		\left| T_{\omega l}(\omega) \right|^2 = \Gamma_{\omega l}(\omega).
	\end{equation}
	The absorption cross-section of the black hole can be expressed as:
	\begin{equation}
		\sigma_{\mathrm{abs}}(\omega) = \sum_{l=0}^{\infty} \sigma_{\mathrm{abs}}^l(\omega) = \frac{\pi}{\omega^2} \sum_{l=0}^{\infty} (2l + 1) \, \Gamma_{\omega l}(\omega),
	\end{equation}
	where $\Gamma_{\omega l}$ is the grey-body factor. The partial absorption cross-section is given by:
	\begin{equation}
		\sigma_{\mathrm{abs}}^l(\omega) = \frac{\pi}{\omega^2} (2l + 1) \, \Gamma_{\omega l}(\omega).
	\end{equation}
	
	\begin{figure*}
		\includegraphics[width=0.8\textwidth]{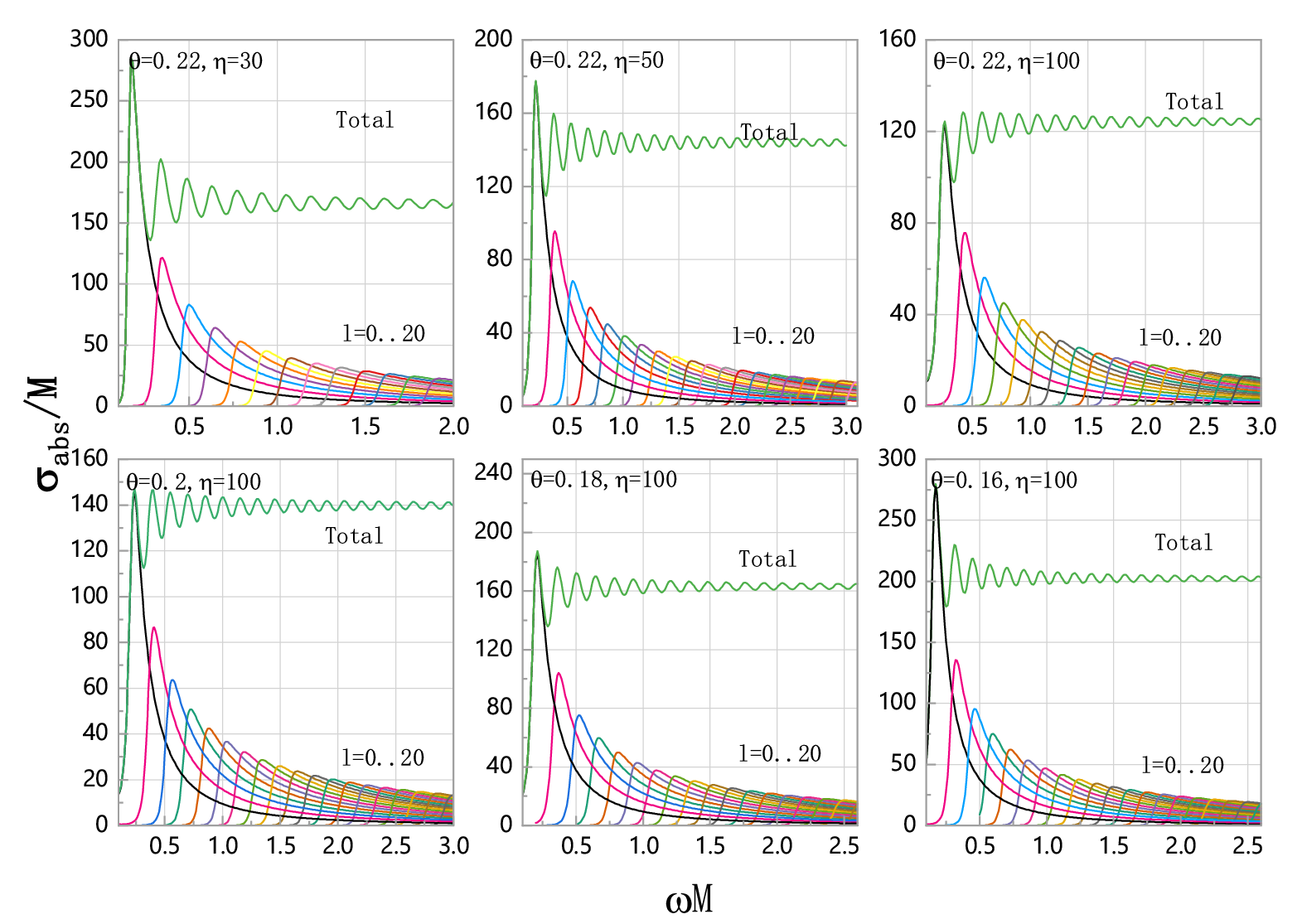}
		\caption{\label{fig5}Partial and total absorption cross-sections for different parameters.}
	\end{figure*}

	Figure \ref{fig5} displays the partial and total absorption cross-sections of the non-commutative black hole for different values of the parameters $\theta_{NC}$ and $\eta$. It can be seen that the partial absorption cross-section decreases gradually with increasing frequency $\omega$. In the low-frequency regime, partial waves with smaller angular quantum numbers $l$ dominate the contribution towards the absorption cross-section. Each oscillation in the total absorption cross-section corresponds to a peak in its partial counterpart. In the high-frequency region, the total cross-section exhibits oscillatory characteristics around the geometric optics limit \cite{sanchez1978absorption}. Moreover, as the angular quantum number $l$ increases, the peak value of the partial absorption cross-section decreases, and the corresponding peak shifts to higher frequencies. This trend is consistent with the variation characteristics of the grey-body factors described previously.
	
	A comparison among different values of $\theta_{NC}$ and $\eta$ shows that both parameters influence the absorption cross-section in a similar way, consistent with their respective effects on the effective potential. For a fixed angular quantum number $l$, smaller values of $\theta_{NC}$ or $\eta$ yield a higher peak in the partial absorption cross-section, and the peak position shifts towards lower frequencies. Meanwhile, the oscillations in the total absorption cross-section become more pronounced, with the oscillation peaks also shifting to the lower-frequency region. Conversely, larger values of $\theta_{NC}$ or $\eta$ result in an overall suppression of the absorption cross-section.
	
		\begin{figure*}
		\includegraphics[width=0.6\textwidth]{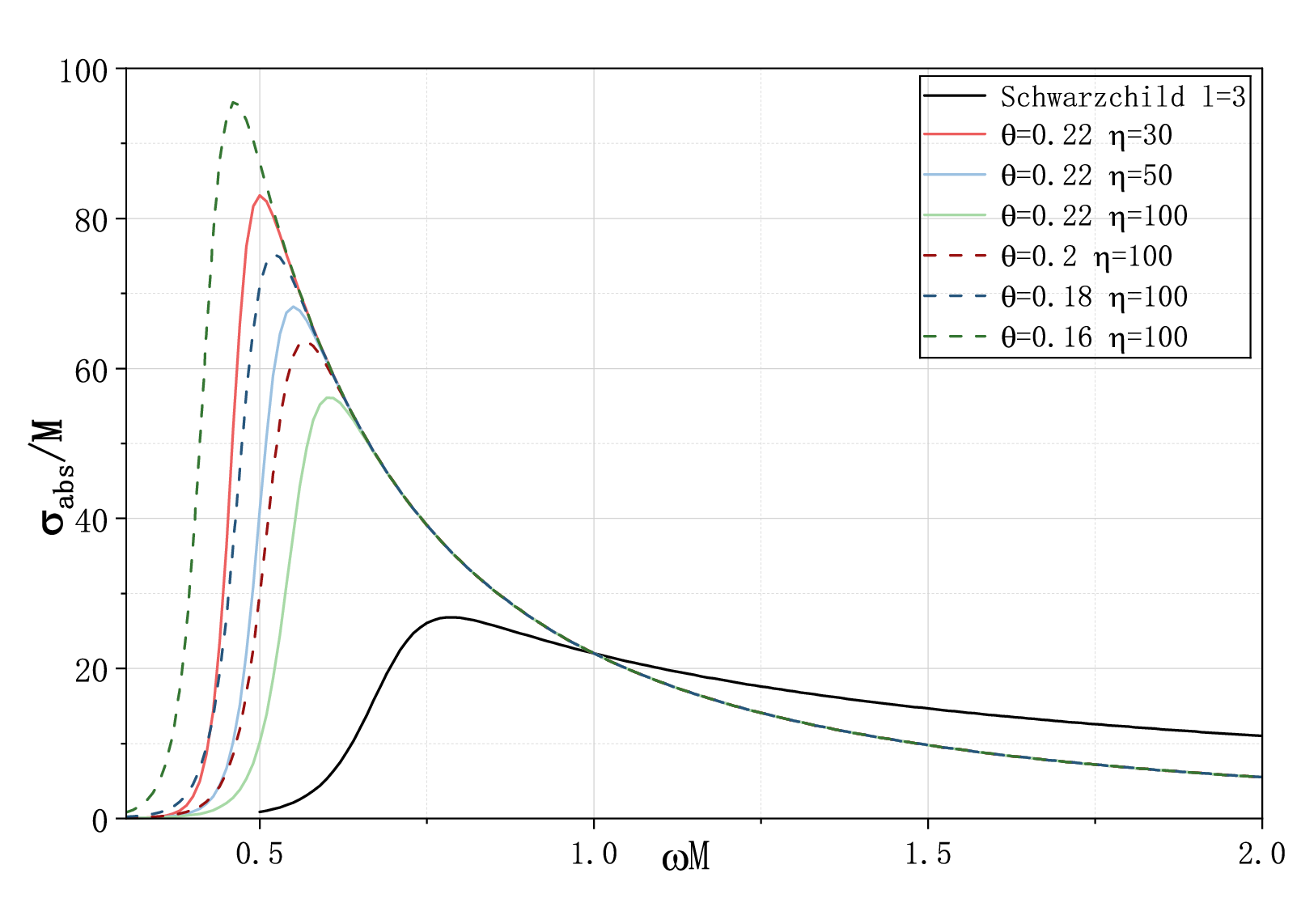}
		\caption{\label{fig6}Partial absorption cross sections of the non-commutative black hole and the Schwarzschild black hole for $l=3$ under different parameter choices.}
	\end{figure*}

	{\color{red} Figure \ref{fig6}  illustrates the partial absorption cross sections for the non-commutative black hole in comparison with the Schwarzschild case for the multipole number $l=3$ under different parameter choices. It is observed that the non-commutative parameter $\theta$ and the coupling constant $\eta$ significantly enhance the absorption capability of the black hole. In particular, the peak values of the absorption cross section are noticeably larger than those of the Schwarzschild black hole for all considered parameter sets.}
	
	\section{Conclusions} \label{sec5}
	In summary, this work investigates the grey-body factors and absorption cross-sections for a scalar field nonminimally coupled to the Einstein tensor in the background of a non-commutative black hole. We analyzed the dependence of these quantities on the frequency and on the parameters $\theta_{NC}$ and $\eta$. {\color{red}Furthermore, we employed the recently established correspondence between grey-body factors and quasinormal modes \cite{konoplya2024correspondence} as a theoretical framework to analyze the scattering properties of non-commutative black holes.The correspondence reproduces the expected behavior with high accuracy in the eikonal regime of large $\ell$, while deviations naturally appear for low angular momentum numbers where WKB-based approaches are known to be less reliable.} 
	
	Our analysis reveals that although the parameters $\theta_{NC}$ (which characterize the fundamental length scale of spacetime non-commutativity) and $\eta$ (which govern the strength of the coupling to the Einstein tensor) originate from distinct physical concepts, their impact on the scattering processes exhibits a remarkable consistency. {\color{red}Our semi-analytical results show that} increasing either parameter shifts the grey-body factor towards lower frequencies and makes its variation more gradual, while modifying both the magnitude and spectral distribution of the absorption cross section and reducing the oscillation amplitude in the total cross-section. These phenomena collectively suggest a core physical picture: both the intrinsic quantum fuzziness of spacetime (encoded in $\theta_{NC}$) and the strengthened gravitational coupling strength (encoded in $\eta$) effectively increase the height and width of the effective potential barrier. This enhancement increases the reflectivity of the barrier and consequently suppresses the absorption capability of the black hole. Thus, despite their different origins, both parameters influence the black hole's interaction with quantum fields in a similar way, primarily by modifying the effective potential landscape.{\color{red}These findings indicate that non-commutativity and Einstein-tensor coupling deform the effective scattering potential while preserving the underlying QNM–greybody correspondence structure, suggesting a universal geometric control of wave scattering even in quantum-corrected spacetimes.}
	
	{\color{red} Our calculations of the grey-body factors and absorption cross-sections reveal that the scattering properties of the scalar field are predominantly governed by the near-horizon geometry of the spacetime. The transmission probability is strongly suppressed in the low-frequency regime and asymptotes to unity in the high-frequency limit, fully consistent with the canonical scattering behavior of black hole spacetimes. }
	
	{\color{red} From an astrophysical perspective, these findings carry important implications for gravitational-wave astronomy. The grey-body factors and absorption cross sections directly encode the near-horizon signatures of black holes, which are imprinted on the gravitational-wave emission and quasi-normal mode ringdown signals detected by LIGO/Virgo/KAGRA observatories \cite{abbott2021tests}. In particular, the low-multipole modes,which dominate astrophysical merger signals, are highly sensitive to modifications of the effective potential induced by spacetime noncommutativity and nonminimal coupling, providing a promising channel to probe new physics beyond general relativity \cite{konoplya2016gw,zhao2023ncqnm}.  Future high-precision gravitational-wave observations may thus distinguish noncommutative black holes from general relativistic Schwarzschild black holes, offering a unique window to probe quantum gravity effects in the strong-field regime. }

	\bibliography{ref.bib}
	
\end{document}